\documentclass[showpacs,pra,superscriptaddress,amsmath,%
amsfonts,amssymb,a4paper,twocolumn]{revtex4}

\usepackage[english]{babel}

\usepackage{amsthm}

\DeclareMathOperator{\im}{Im}

\newtheorem{Th}{Theorem}

\newtheorem{cor}{Corollary}

\usepackage{euscript,amssymb,amsbsy}

\newcommand{\txi}{\widetilde\xi}
\newcommand{\tXi}{\widetilde\Xi}

\newcommand{\diag}{\mbox{\rm diag}\,}

\newcommand{\be}[1]{\begin{equation}\label{#1}}
\newcommand{\ee}{\end{equation}}
\newcommand{\ba}[1]{\begin{eqnarray}\label{#1}}
\newcommand{\ea}{\end{eqnarray}}
\newcommand{\rf}[1]{(\ref{#1})}
\newcommand{\nn}{\nonumber}

\newcommand{\cH}{\mathcal{H}}

\newcommand{\cU}{\mathcal{U}}

\def\a{\alpha}
\def\b{\beta}
\def\d{\delta}

\def\e{\varepsilon}

\def\lb{\lambda}
\def\m{\mu}

\def\O{\Omega}

\def\vfi{\varphi}

\def\ra{\rangle}
\def\la{\langle}
\def\dg{\dagger}

\newcommand{\wt}{\widetilde}

\begin{document}

\title{Optimal positive-operator-valued measures
 for unambiguous state discrimination}

\author{Boris F Samsonov}
\affiliation{
 Physics Department, Tomsk State University, 36 Lenin Avenue,
634050 Tomsk, Russia}

\begin{abstract}
Optimization of the mean efficiency for unambiguous (or error free)
 discrimination
among $N$ given linearly independent nonorthogonal states should be
realized in a way to keep the probabilistic quantum
mechanical interpretation.
This imposes a condition on a certain matrix to be
positive semidefinite.
We reformulated this condition in such a way that the
conditioned optimization problem
for the mean efficiency was reduced to finding an
unconditioned maximum of a function defined on a unit
$N$-sphere for equiprobable states and on an $N$-ellipsoid if
the states are given with different probabilities.
We established that for equiprobable states a point on the
sphere with equal values of Cartesian coordinates,
which we call symmetric point,
 plays a special role.
Sufficient conditions
for a vector set are formulated
for which
the mean efficiency
for equiprobable states takes its
maximal value at the symmetric point.
This set, in particular,
includes previously studied symmetric states.
A subset of symmetric states, for which the optimal
measurement corresponds to a POVM requiring
a one-dimensional ancilla space
is constructed.
 We presented our constructions of a POVM suitable
 for the ancilla space dimension varying from $1$ till $N$ and
 the Neumark's extension differing from the existing
 schemes by the property that it is straightforwardly
 applicable to the case when it is desirable to present
  the whole space system + ancilla
 as the tensor product of a two-dimensional ancilla space and
 the $N$-dimensional system space.
\end{abstract}

\maketitle

\section{Introduction}

Being important for quantum communication, quantum
cryptography and quantum algorithms (see e.g. \cite{1})
the problem of finding an optimal strategy
for discriminating among different
non-orthogonal quantum states has
gained renewed interest \cite{2,3,4,5,6,7,8}
(for reviews see \cite{3,4}).
Between different strategies based on various criteria
(see e. g. \cite{4}) a special role is played by the one called
{\em unambiguous state discrimination}
 \cite{2,3,4,5,6,7,8}. It is based
on pioneering works of Ivanovic, Dieks and Peres \cite{5} who
considered the problem of discriminating among two
non-orthogonal states.
In \cite{6} Peres and Terno examined
the case of three states. In
\cite{7} Chefles
showed that non-orthogonal states may be unambiguously
discriminated if and only if they are linearly independent
and proved an important fact that the
necessary (but not sufficient) condition for a
 measurement to be optimal
consists in equating to unity the
 highest eigenvalue of a certain matrix.
Chefles and Barnett \cite{8} using this property
 derived an upper bound
(achievable by an optimal measurement)
 for success probability
to discriminate among $N$ so called symmetric
states given with equal a priori probabilities.

Another important general result was established by Duan
and
Guo \cite{DG} who derived a matrix inequality for
efficiencies to unambiguously discriminate among $N$ given
linearly independent states. This result permitted to Sun
et al \cite{SZFY}
to reformulate the problem of finding the optimal
discriminating
strategy as the well-known semidefinite
(conditioned) programming problem.
 Although this fact together with the remark that there
 exist softwares for numerical solution of the
 latter problem \cite{SZFY,V} reduces any particular
 optimization task to a technical problem, this approach,
 as well as any numerical solution, usually
 does not bring any additional analytic insight.

In the current paper we show that the conditioned
optimization problem for the mean efficiency
may be reformulated as a problem of
finding an unconditioned maximum of a function
defined on an
$N$-sphere for equiprobable states and on an $N$-ellipsoid
when the states are given with different probabilities
(Theorem \ref{Th1} and Corollary \ref{Cr1}).
We establish that
for equiprobable states
there exists a special point on the unit
$N$-sphere called {\em a symmetric point} where the mean
efficiency takes its maximal value for certain sets of states.
In particular, this set includes all symmetric state.
It happens that
for $N=2,3$ this point is optimal only for symmetric
states.
Nevertheless, as we will show,
 starting from $N=4$
the set of states, for which the symmetric point
is optimal, is wider than
the set of symmetric states.
 Using Theorem \ref{Th1} we formulated
 conditions
 on the set of states which are
  sufficient
for the mean efficiency to be maximal at the symmetric
point (Theorem \ref{th3}).
For $N=4$
Theorem \ref{th3} is illustrated by
an example of a three parameter set of states
which are non-symmetric and
such
that one of the parameters is just the optimal mean
efficiency.

We show that if
one of the eigenvalues of
matrix $\Psi\Psi^\dg$, with $\Psi$ being an $N\times N$
matrix where the given vectors are collected as columns,
is $N-1$ fold degenerate, the set is either a particular
set of symmetric states or their unitary rotated form.
This set may be considered as the simplest generalization
of two states since the expression for the
optimal mean efficiency
for two states is a particular case of a more general
formula obtained for the above set.
Moreover, if this is the lowest eigenvalue, which is $N-1$
fold degenerate, the
ancilla space for the optimal measurement becomes one-dimensional.

For a general set of symmetric states we obtain
for the optimal mean efficiency
an expression
alternative but equivalent to
that found by Chefles and Barnett \cite{8}.

   Finally, we present our constructions of a POVM suitable
 for the ancilla space dimension varying from $1$ till $N$ and
 the Neumark's extension differing from the existing
 schemes by the property that it is straightforwardly
 applicable to the case when it is desirable to present
  the whole space system + ancilla
 as the tensor product of a two-dimensional ancilla space and
 the $N$-dimensional system space.

\section{Optimal POVM\label{OP}}

Assume that we are given a set of non-orthogonal but
linearly independent and
 normalized
to unity vectors (states) $\psi_j$,
$\la\psi_j|\psi_j\ra=1$, $j=1,\ldots,N$,
which we collect as columns to an $N\times N$ matrix
$\Psi=(\psi_1,\ldots,\psi_N)$.
Because of the linear independence of the vectors $\psi_j$
the
matrix $\Psi$ is non-singular, $\mbox{det}\Psi\ne0$,
which is
the property we will assume to hold throughout
the paper.
Below we will frequently use the following terminology. Instead of
saying the vectors (states) $\psi_1,\ldots,\psi_N$ we will say the
vectors (states) $\Psi$.
We will assume also that these vectors span an $N$ dimensional
complex Hilbert space $\cH$ with the usually defined inner
product $\la\cdot|\cdot\ra$.
Evidently, they form a
non-orthogonal basis in $\cH$.
As it was
previously shown \cite{2,3,4,5,6,7,8} unambiguous
(or error free)
strategy
for discriminating among these states, is possible if and
only if they are discriminated by a generalized measurement %
\footnote{Projective (i.e. von Neumann) measurements are particular
cases
of generalized measurements. Therefore they are not excluded
from our consideration.}
(for reviews see e.g. \cite{GM}).
To construct a positive operator valued measure (POVM)
to unambiguously discriminate among the states $\Psi$
it is
convenient to introduce in $\cH$ the basis
$\xi_k$ biorthogonal to $\psi_j$
(Chefles and Barnett  \cite{8} call  $\xi_k$ reciprocal states)
defined as
$\la\xi_k|\psi_j\ra=\d_{kj}$.
Evidently, being collected as columns to matrix
$\Xi=(\xi_1,\ldots,\xi_N)$
they
may be found as
$\Xi=(\Psi^\dg)^{-1}$
where $\Psi^\dg$ is a matrix Hermitian conjugate to $\Psi$.
Note that multiplication of $\Xi$ from the right by a
diagonal matrix
\be{Xd}
X=\mbox{diag}(x_1,\ldots,x_N)
\ee
 corresponds
to changing the norms of the vectors $\xi_j$,
$\xi_{x,j}=x_j\xi_j$,
$\Xi_x=(\xi_{x,1},\ldots,\xi_{x,N})$,
\be{}
\Xi_x=\Xi X=(\Psi^\dg)^{-1}X\,,
\ee
leaving unchanged the orthogonality between $\xi_{xk}$ and
$\psi_j$ for $j\ne k$
\be{ort}
\la\psi_j|\xi_{xk}\ra=x_j\d_{jk}\,.
\ee
By this reason without losing generality one may always assume
 $x_j\ge0$, $j=1,\ldots,N$,
 the property we assume to hold throughout the paper.
 We shall call $X$ the weight matrix and the $x_j$ weights.

Assume now we can associate detection operators with
$\Pi_{xj}=|\xi_{xj}\ra\la\xi_{xj}|$.
Note that the matrix
\be{SS}
\Pi_x\equiv\sum_{j=1}^N|\xi_{xj}\ra\la\xi_{xj}|=\Xi_x\Xi_x^\dg\,,
\ee
which is the left modulus of $\Xi_x$
(see \cite{Gant} for the definition),
is Hermitian and positive definite.
Then because of orthogonality \rf{ort} detector $\Pi_{xk}$
clicks if and only if the state $\psi_k$ is captured.
This
confirms that we can unambiguously discriminate  $\psi_k$
 from all other states $\psi_j$, $j\ne k$.
The probability $P_k$ for this detector to click is
$P_k=\la\psi_k|\Pi_{xk}|\psi_k\ra$
which in view of \rf{ort} reduces to
$P_k=x_k^2$.

This probabilistic interpretation is valid if and only if
there exists a Hermitian positive definite
(or semidefinite) matrix $\wt\Pi_x$ such that
\be{POVMx}
\Pi_x+\wt\Pi_x=I_N
\ee
where $I_N$ is the $N\times N$ identity matrix.
Relation \rf{POVMx} defines a (discrete in the current case)
POVM which is the cornerstone of the probabilistic
interpretation.
From here it follows that
$\wt\Pi_x=I_N-\Pi_x=I_N-\Xi_x\Xi_x^\dg$ should be positive definite (or
semidefinite).
Note that this matrix is positive definite (semidefinite)
together with
$I_N-\Xi_x^\dg\Xi_x=I_N-X(\Psi^\dg\Psi)^{-1}X$ which,
in turn, is positive definite (semidefinite) together with
$\Psi^\dg\Psi-X^2$.
The last condition is just the same necessary and
sufficient condition for an optimal POVM to exist first
found by Duan and Guo \cite{DG}.
We, thus, identify
the square of our weight matrix
 $X^2$ with the efficiency matrix
$\Gamma$ used by Duan and Guo.

Note that $\wt\Pi_x=I_N-\Pi_x$ is positive semidefinite if
the highest eigenvalue of $\Pi_x$ is equal to $1$.
Thus, a POVM can always be constructed if
$X=\lb_m^{-1}I_N$, $(\lb_m>0)$ where $\lb_m^{-2}$
is the highest eigenvalue of
$\Xi\Xi^\dg$
and $\lb_m^2$
 is the lowest eigenvalue
of both
the right modulus of $\Psi$,
$\Psi^\dg\Psi=(\Xi^\dg\Xi)^{-1}$,
 and its left modulus  $\Psi\Psi^\dg$.
 Another important property we would like to emphasize is the
 following.
 When the mean efficiency takes its maximal value,
the highest eigenvalue of $\Pi_x$ is necessarily equal to 1 \cite{8}.
Therefore,  $\wt\Pi_x$ is always positive semidefinite if
the POVM is optimal.
As we will show below in a number of cases this takes place
for $X=\lb_m^{-1}I_N$.
Moreover, $\mbox{\rm rank}\,\wt\Pi_x=N-N_m:=N_a$
where $N_m=1,\ldots,N-1$ is the degree of degeneracy of the highest
eigenvalue of $\Pi_x$
(for $X=\lb_m^{-1}I_N$ this is also the degree of degeneracy of
$\lb_m^{-2}$) so that $N_a=1,\ldots,N-1$.

If the state $\psi_j$ can be chosen from the set $\Psi$ with
a probability $\eta_j$, the total probability  of correctly
identifying any state
\be{px}
P(x)=\sum_{j=1}^N\eta_jP_j=\sum_{j=1}^N\eta_jx_j^2
\ee
is called the mean efficiency.
The optimization problem becomes now a semidefinite programming
problem (see \cite{SZFY}) consisting in finding the maximum of $P(x)$ with
respect to $X\ge0$ subject to either
$\Psi^\dg\Psi-X^2\ge0$ or equivalently
\be{PDC}
I_N-\Xi_x\Xi_x^\dg=I_N-(\Psi^\dg)^{-1}X^2\Psi^{-1}\ge0\,.
\ee
We shall denote this maximum by $P_M$.
In contrast to approach proposed in \cite{SZFY}
the constraint \rf{PDC}, we are using,
 permits us to reduce the
conditioned optimization problem to finding the minimal value of a function
defined on an $N$-ellipsoid.
For brevity in what follows we shall denote by
 $x=(x_1,\ldots,x_N)$ Cartesian coordinates of a point
 from $\mathbb R^{N}$
 and by $t=(t_1,\ldots,t_{N-1})$
curvilinear coordinates on either a unit $N$-sphere or
an $N$-ellipsoid with fixed axes.

\begin{Th}\label{Th1}
Let us denote
\be{}
P_M=\max_{\stackrel{X}{\,
(I_N-\Xi_x\Xi_x^\dg\ge0)}}\sum_{j=1}^N\eta_jx_j^2
\ee
and
\be{la}
\m_M^2=\min_{t\in\O}\m_m^2(t)
\ee
where $\m_m^2(t)$ is the highest eigenvalue of the matrix
$\Xi Y^2(t)\Xi^\dg$ with $Y(t)$ being a diagonal matrix
with the non-zero entries equal coordinates of a point
lying
 on the positive part of
an $N$-ellipsoid with the axes
$\eta_1^{-1/2},\ldots,\eta_N^{-1/2}$.
We denote $\O$ this part of the ellipsoid,
i.e.
\be{Yt}
Y(t)=\diag(y_1(t),\ldots,y_N(t))\,,
\ee
\ba{xx}
y_1(t)&=&\eta_1^{-1/2}\sin t_1\sin t_2\ldots\sin t_{N-2}\sin t_{N-1}\nn \\
y_2(t)&=&\eta_2^{-1/2}\sin t_1\sin t_2\ldots\sin t_{N-2}\cos t_{N-1}\nn \\
y_3(t)&=&\eta_3^{-1/2}\sin t_1\sin t_2\ldots\sin t_{N-3}\cos t_{N-2}\nn \\
&\cdots&\nn \\
y_{N-1}(t)&=&\eta_{N-1}^{-1/2}\sin t_1\cos t_2 \quad (\mbox{\rm for }N>2) \nn \\
y_{N}(t)&=&\eta_{N}^{-1/2}\cos t_1 \\
& & 0<t_1,\ldots,t_N<\pi/2\,. \label{yc}
\ea
Then
\be{PM}
P_M=\m_M^{-2}\,.
\ee
\end{Th}

\begin{proof}
First we note that $\sum_{j=1}^N\eta_jy_j^2=1$.

Let us put in \rf{px} $X=rY(t)$, i.e. $x_j=ry_j(t)$,
$j=1,\ldots,N$.
This yields $P(x)=r^2$
and
$\Xi X^2\Xi^\dg=r^2\Xi Y^2(t)\Xi^\dg$.
Since $r^2$ is simply a scaling factor between
eigenvalues of $\Xi X^2\Xi^\dg$ and $\Xi Y^2(t)\Xi^\dg$
this implies
\be{Px}
\m_m^2=r^{-2}\wt\m_m^2=P^{-1}(x)\wt\m_m^2
\ee
where $\wt\m_m^2$ is the highest eigenvalue of
 $\Xi X^2\Xi^\dg$.
 We note that the left hand side of Eq. \rf{Px}
 is a function defined on the ellipsoid,
 $\m_m=\m_m(t)$.
Next, we shall use the property proven
in \cite{7}:
The necessary condition
for the efficiency to be optimal is
 the highest
eigenvalue of $\Xi X^2\Xi^\dg$ being equal to $1$.
(Under this condition the constraint
$I_N-\Xi_x\Xi_x^\dg\ge0$ is automatically satisfied.)
Let us put in \rf{Px} $\wt\m_m^2=1$ $\forall t\in\O$.
Under this condition
Eq. \rf{Px} defines parameter $r$ and with it the mean
efficiency
$P(x)$ as functions on the ellipsoid
only, i.e. $P=P(t)=r^2(t)=\m_m^{-2}(t)$.
Thus, the maximal value of $P(t)$ corresponds to the minimal value of
$\m_m^2(t)$ on the part of the
ellipsoid with all $y_i(t)\ge0$.
\end{proof}

For equiprobable states, when
$\eta_1=\ldots=\eta_N=1/N$, we find it convenient to re-scale
all $y_j(t)$ and use $\wt y_j(t)=\sqrt{N}y_j(t)$ so that
the point $(\wt y_1(t),\ldots,\wt y_N(t))$ lays on
a unit $N$-sphere.
To simplify notations below in this case we
will use just the unit sphere and omit tilde
over $y_j(t)$.
This leads to the following modification of
Theorem \ref{Th1}.
\begin{cor}\label{Cr1}
For equiprobable $\psi_j$, i.e. for
$\eta_1=\eta_2=\ldots=\eta_N=N^{-1}$,
\be{PMS}
P_M=N^{-1}\max_{t\in\O}\m^{-2}_m(t)
\ee
where $\O$ is a part of the unit $N$-sphere defined by \rf{xx}
and \rf{yc}
with $\eta_1=\ldots=\eta_N=1$, $\m_m^2(t)$ is
the highest eigenvalue of the matrix $\Xi Y^2(t)\Xi^\dg$,
and $Y(t)$ is given in \rf{Yt}.
\end{cor}

We would like to stress that the above condition
for the highest eigenvalue of $\Xi X^2\Xi^\dg$  be equal
to $1$
 proven in
\cite{7} is necessary but not sufficient
for the efficiency to take the maximal value.
As it is clear from the proof of
Theorem \ref{Th1},
for $P(t)=\m_m^{-2}(t)$
this necessary condition
 is satisfied $\forall t\in\O$
and, in general,
$P(t)$ is not constant for $t\in\O$, which can be easily
seen from the simplest $2\times2$ example.
From \rf{Px} it follows also that, in general,
$P(x)=\wt\m_m^{\,2}\m_m^{-2}$ depends on both a point $t$ on
the ellipsoid and parameter $r$.

On the ellipsoid there always exists a special point
$t_0$ where $y_1(t_0)=y_2(t_0)=\ldots=y_N(t_0)$.
We will call this point {the \em symmetric} point.
At this point matrix $Y^2(t)|_{t=t_0}$ becomes proportional to
the identity matrix and $\Xi X^2\Xi^\dg$ is proportional to
$\Xi\Xi^\dg$.
By this reason the eigenvalues of
$\Xi X^2\Xi^\dg$ are, simply, the eigenvalues of $\Xi\Xi^\dg$
scaled by a factor $r^2y_1^2(t_0)$.
In particular,
\be{mm2}
\wt\m_m^2=r^2y_1^2(t_0)\nu_m^2\,,
\ee
where
$\nu_m^2$ is the highest eigenvalue of $\Xi\Xi^\dg$
which is given through the matrix $\Psi$ since
$\Xi=(\Psi^\dg)^{-1}$.
If now we require for the highest eigenvalue of
$\Xi X^2\Xi^\dg$ be equal to 1,
$\wt\m_m^2=1$,
 Eq. \rf{mm2} defines
$r^2$ as a function of $t_0$.
Therefore from Eqs. \rf{Px} and \rf{mm2}
 we obtain the mean efficiency
$P(t_0)=r^2(t_0)=y_1^{-2}(t_0)\nu_m^{-2}$.
For equiprobable states, as we show below,
in a number of cases point $t_0$ is just
the maximal point for the mean efficiency.
In particular, this point is optimal for any symmetric
set of equiprobable states
but not only for these states.
The set of states for which the mean
efficiency takes its maximal value at $t=t_0$ is wider than
the set of symmetric states.
However,
as our numerical tests show, in many cases there exist
points
on the ellipsoid where the efficiency takes higher values.
We believe that the symmetric point
 may be a good starting point for a numerical
optimization.

Theorem \ref{Th1} and Corollary \ref{Cr1} give a general
recipe for finding the optimal
weight matrix $X(t)$ and with it the optimal POVM defined by
the
set $\Xi_x=\Xi X(t)$ to
unambiguously discriminate among the vectors collected to matrix
$\Psi$.
In general the matrix $\Psi$ is non-Hermitian and can have complex
eigenvalues.
Nevertheless, as we show below, for any given $\Psi$ there
always exists a positive definite $\Psi_1=\Psi_1^\dg$ which
has the same optimal weight matrix and the same optimal
mean efficiency as $\Psi$.
\begin{cor}\label{prop}
Both
the optimal weight matrix $X=X_M$ \rf{Xd}
and the optimal mean efficiency $P_M$
 are invariant
under a unitary rotation of the vector set $\Psi$.
In particular,
for any non-singular $\Psi$ given in the form
\ba{Psi}
\Psi=U_2E_0U_1\,, & U_{1,2}^{-1}=U_{1,2}^\dg\,,
 \\
E_0=\mbox{diag}(E_{0,1},\ldots,E_{0,N})\,,
&\! E_{0,j}>0, j=1,\ldots,N
\label{PsiE}
\ea
the optimization may be realized for
 $\Psi_1=U_1^\dg E_0 U_1$.
\end{cor}
\begin{proof}
First we note that
 decomposition \rf{Psi}, \rf{PsiE} takes place for
 any non-singular matrix $\Psi$ (see e.g. \cite{Gant}).
 Denote $t=t_M$ the point on the ellipsoid
where the highest
 eigenvalue $\m_m^2(t)$ of the matrix $\Xi Y^2(t)\Xi^\dg$
takes its minimal value as a function of $t$.
Next,
since
 $\Xi=(\Psi^\dg)^{-1}$,
this implies
$\Xi Y^2(t)\Xi^\dg=U_2E_0^{-1}U_1Y^2(t)U_1^\dg E_0^{-1}U_2^\dg$.
From here it follows that the eigenvalues of matrix
$\Xi Y^2(t_M)\Xi^\dg$
 do not depend on unitary matrix $U_2$.
In particular,
the eigenvalue $\m_M^2=\m_m^2(t_M)$ \rf{la}
and with it,
as it follows from \rf{PM},
both
 the optimal mean efficiency $P_M$ and
 the weight matrix $X_M=\sqrt{P_M}Y(t_M)$
are
invariant under a unitary rotation of the vectors $\Psi$.
Therefore the optimization may be realized
for the set
  $\Psi_1=U_1^\dg E_0 U_1$.
\end{proof}

Note that in general properties
of the vectors collected to a matrix
unitarily equivalent to a given matrix change essentially.
In particular, if the vectors $\Psi$ are normalized to
unity, the vectors $U\Psi U^\dg$, in general, are not
normalized to unity.
Nevertheless, there exist
particular unitary transformations keeping unchanged both
the mean efficiency and the weight matrix. These are
transformations from an Abelian subgroup of the unitary
group.
They have the form
$U=U_a=\mbox{diag}(e^{i\b_1},\ldots,e^{i\b_N})$, $\b_j\in\mathbb R$.
Indeed, being multiplied from the right by $U_a$
every column vector of $\Psi$ acquires
 an additional inessential phase factor
 whereas after
the left multiplication the vectors
$\Psi$ are simply unitary rotated.
Thus, such a transformation corresponds to a re-scaling of the
vector set $\Psi$ by phase factors
followed by their unitary rotation.
According to Corollary \ref{prop} these operations can
affect neither the mean efficiency nor the weight matrix.


The next theorem establishes
sufficient conditions
for the symmetric point on the $N$-sphere to be a point of the
maximal mean efficiency for equiprobable states.
To prove this result,
according to Corollary \ref{Cr1} one has to find
the minimum of the
highest eigenvalue of the matrix $\Xi Y^2(t)\Xi^\dg$ for $t$ on
the unit $N$-sphere.

\begin{Th}\label{th3}
Let $\Psi$ be Hermitian, positive definite
and the states $\Psi$ are equiprobable.
If there exists an eigenvector
$\vfi_m^0=(\vfi_{m,1}^0,\ldots,\vfi_{m,N}^0)^t$,
$\la\vfi_m^0|\vfi_m^0\ra=1$,
 of $\Psi$
corresponding to its lowest
 eigenvalue $\lb_m$, i.e.
$\Psi\vfi_j^0=\lb_j\vfi_j^0$, $\lb_m=min_j\lb_j$,
such that
$|\vfi_{m,1}^0|=|\vfi_{m,2}^0|=\ldots=|\vfi_{m,N}^0|=N^{-1/2}$,
then
$P_M=\lb_m^{2}$.
\end{Th}

\begin{proof}
First we note that because of Hermiticity of $\Psi$,
matrices $\Psi$,
 $\Psi\Psi^\dg=\Psi^\dg\Psi=\Psi^2$,
 $\Xi=\Psi^{-1}$ and $\Xi^\dg\Xi=\Xi\Xi^\dg=\Xi^2$
may have the same set of eigenvectors.
 In particular,
 $\Xi^\dg\vfi_m^0=\lb_m^{-1}\vfi_m^0$ and
$\Xi\Xi^\dg\vfi_m^0=\lb_m^{-2}\vfi_m^0$ so that $\lb_m^{-2}$
is the highest eigenvalue of $\Xi\Xi^\dg$.

Let us denote $H(t)=\Xi Y^2(t)\Xi^\dg$ where
$Y(t)$ is given in \rf{Yt}
with $y_j(t)$
being given in \rf{xx} and \rf{yc} with
$\eta_1=\ldots=\eta_N=1$
so that $\sum_jy_j^2(t)=1$.
Let also $\m_m^2(t)$ be the
highest eigenvalue of $H(t)$ and $\vfi_m(t)$ be one of the
eigenvectors with eigenvalue $\m_m^2(t)$. i.e.
\be{}
H(t)\vfi_m(t)=\m_m^2(t)\vfi_m(t)\,.
\ee
Since $\m_m^2(t)$ is the highest eigenvalue then according to
the variational principle
the following inequality
$\m_m^2(t)\ge\la\vfi|H(t)|\vfi\ra$ holds true
$\forall\vfi\in\cH$.
In particular, for any $t$ we have
\ba{ineq}\nn
\m_m^2(t)\ge\la\vfi_m^0|H(t)|\vfi_m^0\ra=
\la\vfi_m^0|\Xi Y^2(t)\Xi^\dg|\vfi_m^0\ra \\ =
\!\lb_m^{-2}\la\vfi_m^0| Y^2(t)|\vfi_m^0\ra\!=\!
\lb_m^{-2}\sum_{i=1}^N|\vfi_{m,i}^0|^2y_i^2\!=\!
N^{-1}\lb_m^{-2}\,.
\ea

Let $t_0$ be a point on the unit sphere where
$y_1(t_0)=\ldots=y_N(t_0)=N^{-1/2}$ and, hence,
$Y(t)=N^{-1/2}I_N$.
This implies that $\vfi_m(t_0)=\vfi_m^0$
and $\m_m^2(t_0)=\la\vfi_m^0|H(t_0)|\vfi_m^0\ra=
N^{-1}\la\vfi_m^0|\Xi\Xi^\dg|\vfi_m^0\ra=
N^{-1}\lb_m^{-2}$.
Comparing this result with \rf{ineq}
we conclude that $\m_m^{-2}(t_0)\ge \m_m^{-2}(t)$ for any $t$ and, hence,
according to Corollary \ref{Cr1}
$P_M=[N\m_m^2(t_0)]^{-1}=\lb_m^{2}$.
\end{proof}

This theorem has an interesting implication
which we formulate as the next theorem.
\begin{Th}\label{th4}
Let matrix $\Psi$ be such that matrix
$\Psi\Psi^\dg$ has only two distinct eigenvalues
$\lb_1^2$ and $\lb_N^2$ and one of them is
$N-1$ fold degenerate,
 then for equiprobable states
$P_M=\min(\lb_1^{2},\lb_N^2)$.
\end{Th}

We would like to remind that
the positive square roots $\lb_j$ of the eigenvalues
 of the left modulus of
$\Psi$, i.e. of the eigenvalues $\lb_j^2$ of $\Psi\Psi^\dg$, are
called singular numbers of $\Psi$ \cite{Gant}.
Thus, by the theorem
assumption one of these numbers is $N-1$ fold degenerate.

\begin{proof}
Let for definiteness $\lb_1^2$ be $N-1$ fold degenerate.
Using Corollary \ref{prop}
without losing generality one may assume
 $\Psi$ to be Hermitian and positive definite,
 $\Psi=U^\dg E_0U$, $U^\dg=U^{-1}=(u_{k,j})$,
 $E_0=\mbox{diag}(E_{0,j})$,
 $E_{0,j}=\lb_1>0$, $j=1,\ldots,N-1$, $E_{0,N}=\lb_N>0$,
 so that matrix $\Psi^\dg\Psi=\Psi\Psi^\dg=\Psi^2$
 has eigenvectors $U^\dg$.
Thus the first $N-1$ vectors
$u_k=(u_{1,k},\ldots,u_{N,k})^t$, $k=1,\ldots,N-1$
 collected as columns to $U^\dg$ correspond to
the eigenvalue $\lb_1$ of $\Psi$
and the last column of $U^\dg$,
$u_N=(u_{1,N},\dots,u_{N,N})^t$,
corresponds to $E_{0,N}=\lb_N$.
 Moreover,
since any unitary transformation
$\Psi\to U\Psi$, $U^{-1}=U^\dg$
does not affect the norms
of the vectors,
diagonal entries
 of $\Psi^2=\Psi^\dg\Psi$
 remain equal to $1$ after any unitary rotation.
 (Note nevertheless that the whole overlap matrix
 $\Psi^\dg\Psi=\Psi^2$ remain unchanged
 after such a transformation.)
 Below we prove that under these conditions
 the eigenvector $u_N$ of $\Psi^2$ corresponding to $\lb_N^2$ has coordinates
 with equal absolute values.
 Moreover,
in the eigensubspace of $\Psi^2$
corresponding to eigenvalue $\lb_1^2$
 there always
 exists  an eigenvector
$u_{N-1}=(u_{1,N-1},\ldots,u_{N,N-1})^t$,
$\la u_{N-1}|u_{N-1}\ra=1$,
$\Psi^2u_{N-1}=\lb_1u_{N-1}$,
such that
$|u_{1,N-1}|=|u_{2,N-1}|=\ldots=|u_{N,N-1}|=N^{-1/2}$.
Then the statement will follow from Theorem \ref{th3}.

Condition $(\Psi^2)_{k,k}=1$ or explicitly
\be{}
\lb_1^2\sum_{j=1}^{N-1}|u_{k,j}|^2+|u_{k,N}|^2\lb_N^2=1\,,
\quad
k=1,\ldots,N
\ee
and unitarity of $U$,
$\sum_{j=1}^{N-1}|u_{k,j}|^2+|u_{k,N}|^2=1$,
imply
\be{uN2}
|u_{k,N}|^2=\frac{1-\lb_1^2}{\lb_N^2-\lb_1^2}\,,
\quad
k=1,\ldots,N\,.
\ee
Using this result and unitarity of $U$ once again one gets
\be{uN}
u_{k,N}=\frac{1}{\sqrt{N}}\,e^{i\a_{k,N}}\,,\quad
k=1,\ldots,N\,,
\ee
where real phases $\a_{k,N}$ are defined by a concrete
choice of $\Psi$.

From \rf{uN2} and \rf{uN} we
find a link between eigenvalues $\lb_1^2$ and
 $\lb^2_N$
\be{link}
\lb_N^2+(N-1)\lb_1^2=N\,,
\ee
which results in a restriction
for the lowest eigenvalue $0<\min(\lb_1^2,\lb_N^2)<1$.

Since by assumption the first $N-1$ vectors collected to
$U^\dg$ correspond to the same eigenvalue of $\Psi$
there is a
freedom to choose in the eigensubspace spanned by these
vectors any orthonormal basis.
For our purpose it is sufficient to show that
there exist real phases $\a_{k,N-1}$, $k=1,\ldots,N$
such that a vector
$u_{N-1}=(u_{1,N-1},\ldots,u_{N,N-1})^t$,
\be{uk0}
u_{k,N-1}=\frac{1}{\sqrt{N}}\,e^{i\a_{k,N-1}}\,,
\quad k=1,\ldots,N
\ee
is orthogonal to the vector $u_{k,N}$ \rf{uN}.
Indeed, from \rf{uN} and \rf{uk0} one
obtains
\be{SRI}
\sum_{k=1}^Ne^{i(\a_{k,N}-\a_{k,N-1})}=0\,.
\ee
This equation means that $\a_{k,N}-\a_{k,N-1}$,
$k=1,\ldots,N$ is
$(k-1)$st root of unity, i.e.
$\a_{k,N}-\a_{k,N-1}=\frac{2\pi}{N} (k-1)$ so that
\be{roots}
\a_{k,N-1}=\a_{k,N}-\frac{2\pi}{N}(k-1)\,,\quad
k=1,\ldots,N\,.
\ee
We, thus, explicitly constructed the vector
$u_{N-1}$
from the
eigensubspace corresponding to the eigenvalue $\lb_1$
having the necessary form.
\end{proof}

From the proof of this theorem one can extract a particular
representation of a vector set $\Psi$ when $\Psi$ is
Hermitian, positive definite and has only two distinct
eigenvalues.
First we note that using proper re-scaling and unitary
rotation one can always guarantee equal values of all
coordinates of the vector $u_N$, i.e.
$u_N=(1,\ldots,1)^t/\sqrt{N}$.
By this reason the vectors
$u_j=(\exp(i\a_{1,j}),\ldots,\exp(i\a_{N,j}))^t/\sqrt{N}$,
$j=1,\ldots,N-1$
with
$\a_{k,j-1}=\a_{k,j}-2\pi(k-1)/N$,
$k=1,\ldots,N$, $j=N,N-1,\ldots,2$
are orthogonal to $u_N$ since they differ from each other
only by permutations of coordinates.
Moreover, it is
straightforward to check that together with $u_N$ they form
an orthonormal set in $\cH_N$.
(They correspond to all $N$ roots of unity.)
Using property \rf{SRI} of roots of unity
one can find the entries $\psi_{j,k}$ of $\Psi=U^\dg E_0U$
\ba{NN1}
\psi_{j,k}=(\lb_N-\lb_1)/N\,,\quad
 j\ne k\,,
\\
 \psi_{k,k}=[\lb_N+(N-1)\lb_1]/N\,,\quad
j,k=1,\ldots,N \label{NN2}
\ea
 where $\lb_1$ and $\lb_N$ should satisfy
Eq. \rf{link}.

Note that the inner products of the vectors $\Psi$
\rf{NN1}, \rf{NN2} are equal to the same value
$\la\psi_i|\psi_j\ra=s=(\lb_N^2-\lb_1^2)/N$,
$i\ne j$, $|s|<1$.
Using Eq. \rf{link} one can
express eigenvalues of $\Psi^2$ in terms of $s$ as
$\lb_1^2=1-s$, $\lb_N^2=\lb_1^2+Ns$.
In turn, the optimal mean efficiency in terms of $s$ reads
\be{Ps}
P_M=1-s+{N}\left(s-|s|\right)/{2}\,.
\ee
Our last comment here is that
$N_a=\mbox{\rm rank}\,\wt\Pi_x=1$
 for $s>0$ and $N_a=N-1$ for $s<0$.

\section{Particular cases}

In this section we show that
for $N=2$ and $N=3$
the maximal mean efficiency obtained from
Theorem \ref{th3} corresponds to a set of
 symmetric equiprobable states
 and this theorem may lead to a wider set of equiprobable states
 starting from $N=4$.
 In Section \ref{N2_sec} we illustrate our theorems for the
 case of two non-orthogonal states.
 In particular, if the
 states are given with different prior probabilities we
 specify a domain in the parameter space when Theorem
 \ref{Th1} results in a von Neumann measurement.
In the sections \ref{N3}, \ref{N4} and \ref{SymS} we
consider equiprobable states only.
 In Section \ref{N3} we construct a four-parameter
 set of three vectors normalized to unity and described by a
 Hermitian positive definite $3\times 3$ matrix and present an example
 of states for which the symmetric point is not optimal.
 In Section \ref{N4} we construct a
 three parameter set of four normalized to
 unity vectors described by a $4\times 4$ Hermitian matrix
 with the optimal point being the symmetric point.
 In Section \ref{SymS} we show that for any set of $N$ symmetric
 states the symmetric point is the maximal point for the
 mean efficiency and re-derive the result
 obtained by Chefles and
 Barnett \cite{8} for the value of the maximal mean
 efficiency.
 We obtain also an alternative formula for the
 maximal mean efficiency which may be useful when the
 states are given in a form different from that used by
 Chefles and Barnett \cite{8}.

\subsection{$N=2$\label{N2_sec}}

Although the case of two states is well studied
(see e.g. \cite{5,2,8,SZFY}) we find it useful
for illustrating our theorems above.

The most general Hermitian $2\times2$ matrix $\Psi$
composed of the vectors normalized to unity has the form
\be{N2}
\Psi=\left(
\begin{array}{cc}\sqrt{1-r^2} & e^{-i\a}r\\
e^{i\a}r &\sqrt{1-r^2}\end{array}\right),
\quad  0<r<\frac{1}{\sqrt{2}}\,.
\ee
The inequality in \rf{N2} guarantees the positivity of
eigenvalues of $\Psi$, $\lb_{1,2}=\sqrt{1-r^2}\mp r$.

First we note that using a matrix
$U_a=\diag(1,\exp(-i\b))$ for
rescaling and unitarily rotating
the set \rf{N2}, i.e
 $\Psi\to U_a\Psi U_a^\dg$,
  we can eliminate the
inessential phase factors in the off-diagonal entries of
$\Psi$.
Therefore in \rf{N2} we can put $\a=0$.
Now we see
that if
for matrix \rf{N2} with $\a=0$
 we displace the first coordinate of the vector
$\psi_1$ to the next position and put its last
(second in the current case) coordinate at the first place,
we obtain $\psi_2$.
The same transformation applied to
$\psi_2$ gives $\psi_1$.
Since this transformation
is unitary, the states \rf{N2} are symmetric
(see \cite{8} for a definition).

For equiprobable states
using $Y(t)={\rm diag}(\sin t,\cos t)$ we obtain the
highest eigenvalue of $\Xi Y^2(t)\Xi=\Psi^{-1}Y^2(t)\Psi^{-1}$
\be{mu2t}
\mu^2_m(t)=
\frac{2-\sqrt{2}\left[1+4r^2-4r^4+(1-2r^2)^2\cos(4t)\right]^{1/2}}%
{4(1-2r^2)^2}
\ee
which has a minimum
($\mu^{-2}_m(t)$ has a maximum)
at $t=\pi/4$.
Note that this is just the symmetric point on the unit
circle and, hence, for $N=2$ the mean efficiency acquires
the maximal value at the symmetric point.
Thus, using Corollary \ref{Cr1} one obtains the
optimal mean efficiency
$P_M=1-2r\sqrt{1-r^2}=1-|\la\psi_1|\psi_2\ra|$.

The coordinates of
eigenvectors of $\Psi$,
$u_{1,2}=(\mp 1+e^{i\a})^t/\sqrt{2}$
have equal absolute values.
Therefore according to Theorem \ref{th3},
$P_M={\rm min}(\lb_1^2,\lb_2^2)=(\sqrt{1-r^2}-r)^2$
which is just the same as above.

For $N=2$ the degree of degeneracy of the lowest
eigenvalue assumed in Theorem \ref{th4} is $N-1=1$.
Thus using Eq. \rf{Ps} for $N=2$ one obtains the same
value once again
$P_M=1-|s|=1-|\la\psi_1|\psi_2\ra|$,
the result previously reported by numerous authors
\cite{5,2,8,SZFY}.
This means that Eq. \rf{Ps} presents the simplest
generalization from $N=2$ to an arbitrary $N$.

For states given with different probabilities
$\eta_1$ and $\eta_2=1-\eta_1$, $\eta_1\ne1/2$ it is
instructive to illustrate the case when Theorem \ref{Th1}
leads to a projective (i.e. von Neumann) measurement.

First we note that without losing generality one may
assume $1/2<\eta_1<1$.
Then choosing
 $Y(t)={\rm diag}(\eta_1^{-1/2}\sin t,\eta_2^{-1/2}\cos t)$
 we find the highest eigenvalue of $\Xi Y^2(t)\Xi$
 \be{musqr}
\mu_m^2(t)=
\frac{\cos(2t)(2\eta_1-1)-2G(t)}{4\eta_1\eta_2(1-2r^2)^2}
 \ee
where
\be{G}
G(t)=[\textstyle{\frac12}+
(\eta_1-\textstyle{\frac12})\cos(2t)]^2-
\eta_1\eta_2(2r^2-1)^2\sin^2(2t)\,.
\ee
Function \rf{musqr} may have minima only at points
where $d\mu_m^2(t)/dt=0$, i.e. at points satisfying the
equation
\be{eqmin}
\sin(2t)G_1(t)=0
\ee
where
\be{G1}
G_1(t)=4\eta_1-2+\frac{2\eta_1-1+(1-16r^2)(1-r^2)\eta_1\eta_2}{G(t)}\,.
\ee
In particular this may happen at $t=0$ and $t=\pi/2$ where
the function $\mu_m^2(t)$ takes the values
\be{mu0pi}
\begin{array}{l}
\mu_m^2(0)=(1-2r^2)^{-2}\eta_2^{-1}\,,\\
\mu_m^2(\textstyle{\frac{\pi}{2}})=(1-2r^2)^{-2}\eta_1^{-1}<\mu_m^2(0)\,.
\end{array}
\ee
Note that at these points either $x_1=0$ or $x_2=0$
meaning that either detection
operator $\Pi_{x1}$ or $\Pi_{x2}$ does not participate at
the optimal measurement and can be omitted from POVM
so that the resolution of the identity \rf{POVMx} contains
in this case
two terms only.
For one-dimensional operators and $N=2$ this necessarily
leads to orthoprojectors as detection operators and, hence,
to a von Neumann measurement.

According to \rf{musqr} and \rf{G} we have
$\mu_m^2(t)=\mu_m^2(\pi-t)$
and therefore $t=\frac{\pi}{2}$ cannot be an inflexion
point for this function.
Moreover there exists no more than one point
$\tilde t\in (0,\pi/2)$ where
$d\mu_m^2(\tilde t)/dt:=(d\mu_m^2(t)/dt)|_{t=\tilde t}=0$.
This follows from the fact that this point should be found
from the equation $G_1(\tilde t)=0$
with $G_1(t)$ given in \rf{G1}
where, according to \rf{G},
 $G(t)$ is a second order polynomial on $\cos(2t)$.
By this reason to specify a domain in the space of parameters
$(r,\eta_1)$ where the optimal measurement is a von Neumann
measurement it is sufficient to find the values of
$(r,\eta_1)$ such that $t=\frac{\pi}{2}$ is just
the point of a minimum for $\mu_m^2(t)$,
i.e. the point where $d^2\mu_m^2(\pi/2)/dt^2>0$.
Thus from Eqs. \rf{mu2t} and \rf{G} we find the condition
\be{}
2+2(4r^4-4r^2-1)\eta_1>0
\ee
or equivalently
\be{}
\frac{1}{1+4r^2-4r^4}<\eta_1<1\,.
\ee
From Eq. \rf{mu0pi} we find the optimal mean efficiency for
this measurement
$P_M=\eta_1(1-2r^2)^2=\eta_1(1-\la\psi_1|\psi_2\ra^2)$
which is just the ``von Neumann value''
$\eta_1\cos^2(\frac{\pi}{2}-\theta)=
\eta_1\sin^2\theta$ where $\theta$ is the angle between the
vectors $\psi_1$ and $\psi_2$.

Below we will consider equiprobable states only.

\subsection{$N=3$\label{N3}}

The most general Hermitian $3\times3$ matrix
$\Psi=(\psi_1,\psi_2,\psi_3)$,
contains three
real and three complex parameters.
Since the vectors $\psi_j$ are assumed to be normalized to
unity, the diagonal entries of $\Psi$ can always be
expressed in term of off-diagonal elements.
Moreover, one can always re-scale the vectors $\psi_2$
and $\psi_3$ such that their first components become real.
After re-scaling  the Hermiticity of $\Psi$ can be
restored by a proper unitary rotation of
the vectors $\Psi$.
Thus, in the current case
without losing generality
 matrices $\Psi$ may be assumed to form
a four-parameter family with a real parametrization.
In
general, eigenvalues of such a matrix should be found from
a third order algebraic equation.
Although any third order
algebraic equation can be solved analytically,
solutions, in general,
are rather complicated and difficult for analyzing.
Therefore below we show that there exists a
parametrization where both the eigenvalues and eigenvectors
of $\Psi$
have a rather simple form but first we find convenient to
illustrate our Theorem \ref{th3}.

Let $u_j=(u_{1,j},u_{2,j},u_{3,j})^t$, and $\lb_j>0$, $j=1,2,3$
be eigenvectors and eigenvalues of $\Psi$ respectively so
that $\Psi=U_3^\dg E_0U_3$, where $U_3^\dg=(u_1,u_2,u_3)$ and
$E_0={\rm diag}(\lb_1,\lb_2,\lb_3)$.
According to Theorem \ref{th3} the components of (for instance)
 vector $u_3$,
should have equal absolute values.
First we note that since the state vectors are defined up
to phase factors one can always choose
$u_{3,3}=1/\sqrt{3}$.
Then
using a similar
re-scaling and unitary rotation
of the vectors $\Psi$ as discussed above,
i.e. $\Psi\to U_a\Psi U_a^\dg=
U_aU_3^\dg U_a^\dg E_0U_aU_3U_a^\dg$
with
$U_a={\rm diag}[\exp(i\b_j)]$, $j=1,2,3$,
one can always guarantee the
choice $u_{1,3}=u_{2,3}>0$.
Since this transformation does not affect $u_{3,3}$,
from here
 it follows
that $u_{1,3}=u_{2,3}=u_{3,3}=1/\sqrt{3}$.
Since $\Psi$ is Hermitian, its eigenvectors $u_j$ are orthogonal
to each other.
Therefore the
vectors $u_1$ and $u_2$
can be obtained by applying
 a general transformation
(rotation) from the group $SU_2$
to arbitrary two orthonormal vectors
chosen to be orthogonal to the vector $u_3=(1,1,1)^t/\sqrt{3}$.
In particular, one can choose
$u_1=(-2,1,1)^t/\sqrt{6}$ and $u_2=(0,1,-1)^t/\sqrt{2}$.
In this way we
obtain the following unitary matrix
\be{U3}
U_3^\dg=\left(
\begin{array}{lll}
-\sqrt{\frac{2}{3}}K_{z\a}&-i\sqrt{\frac{2}{3}}K^*_{xy}&
\frac{1}{\sqrt{3}}\\[.5em]
\frac{1}{\sqrt{6}}K_{z\a}+\frac{i}{\sqrt{2}}K_{xy} &
\frac{i}{\sqrt{6}}K_{xy}^*+\frac{1}{\sqrt{2}}K_{z\a}^*&
\frac{1}{\sqrt{3}}\\[.5em]
\frac{1}{\sqrt{6}}K_{z\a}-\frac{i}{\sqrt{2}}K_{xy} &
\frac{i}{\sqrt{6}}K_{xy}^*-\frac{1}{\sqrt{2}}K_{z\a}^*&
\frac{1}{\sqrt{3}}
\end{array}
\right)
\ee
where
$K_{z\a}=\cos\frac{\a}{2}+iK_z\sin\frac{\a}{2}$,
$K_{xy}=(K_x+iK_y)\sin\frac{\a}{2}$ and we have used the usual
parametrization for the $SU_2$ group \cite{ED} with
$0\le\a\le2\pi$ and $K_x$, $K_y$, $K_z$ being the
coordinates of a real unit 3-dimensional vector,
$K_x^2+K_y^2+K_z^2=1$, $K_{x,y,z}\in \mathbb R$.
Below we find more convenient to use two complex parameters
 $K_{xy}$ and $K_{z\a}$  subject to the normalization
condition
\be{NCK}
|K_{xy}|^2+|K_{z\a}|^2=1\,.
\ee

The normalization condition for the vectors $\Psi$,
$(\Psi^2)_{j,j}=1$, or explicitly
\be{NC}
\sum_{j=1}^N\lb_j^2|u_{k,j}|^2=1\,,\quad k=1,2,\ldots,N\,,\quad N=3\,,
\ee
may be considered as a system of linear
inhomogeneous equations with respect to eigenvalues
$\lb_j^2$.
Note that because of the unitarity of $U_3$ this
system always has a solution $\lb_1^2=\lb_2^2=\lb_3^2=1$.
This trivial
solution is not suitable for our purposes and we need
another one.
This means that system \rf{NC} should be linearly
dependent with a vanishing main determinant,
${\rm det}(|u_{k,j}|^2)=0$.
In this case one of the
eigenvalues, say $\lb_3^2$, is arbitrary while two others are
linear functions of $\lb_3^2$.
It is straightforward to check that
for $u_{k.j}$ given in
\rf{U3} the determinant vanishes and from the system \rf{NC}
we find
\be{lb12}
\lb_1^2=\lb_2^2=\frac{3-\lb_3^2}{2|K_{xy}|^2+2|K_{z\a}|^2}=
\frac{3-\lb_3^2}{2}\,.
\ee
Since we assume $\Psi$ to be positive definite, this means
that $\lb_1=\lb_2$ and the unitary rotation of the vectors
$u_1$ and $u_2$ considered above does not affect $\Psi$.
Thus the vectors $\Psi$ are parameterized by one parameter
($\lb_3$) only,
\be{Psie3}
\Psi=\left(
\begin{array}{ccc}
A_0 & A_1 & A_1 \\ A_1 & A_0 & A_1 \\ A_1 & A_1 & A_0
\end{array}
\right),
\ee
where
\be{A01}
A_0=\frac13(\lb_1+\lb_2+\lb_3)\,,\quad
A_1=\frac13(\lb_3-\lb_2)\,.
\ee
From here we see that
similarly to the previous section
the coordinates of the vector $\psi_2$
are obtained from the coordinates of the vector $\psi_1$ by a
simple permutation (denote it $P$) and the coordinates of the vector $\psi_3$
are obtained from the coordinates of the vector $\psi_2$ by the
same permutation $P$. Applying $P$ to the vector $\psi_3$
one obtains $\psi_1$. This means that $P^3=I_3$ (identity).
Since such a permutation of coordinates is a
unitary transformation, the set \rf{Psie3} is a particular case
of symmetric states (for a definition see e.g. \cite{8}).
From theorems \ref{th3} and \ref{th4}
for the states \rf{Psie3}
we find the optimal mean efficiency
$P_M=\lb_3^2$ for $0<\lb_3<1$ and
$P_M=\frac{3-\lb_3^2}{2}$ for $1<\lb_3<\sqrt{3}$.
Note that the vectors \rf{Psie3} have equal inner products
$s=\la\psi_j|\psi_k\ra=\frac13(\lb_3^2-\lb_1^2)$, $j\ne k$,
$-1/2<s<1$ and the optimal mean efficiency is obtained
from Eq. \rf{Ps} at $N=3$.

The condition $\lb_1^2\ne\lb_2^2$ is incompatible with the
system \rf{NC} when it is considered as a system of linear
equations with respect to $\lb_j^2$.
Nevertheless, for
$\lb_1^2\ne\lb_2^2$ this system has a solution with respect
to $SU_2$ group parameters.
 In particular, after simple
calculations one finds
\be{Kqp}
|K_{z\a}|^2=\frac12\,,\quad K_{xy}=\pm K_{z\a}
\ee
For this parameter set matrix $\Psi$ assumes the form
\be{Psi2pq}
\Psi=\left(
\begin{array}{ccc}
A_0&A_2&A_2^*\\ A_2^*&A_0&A_2\\ A_2&A_2^*&A_0
\end{array}
\right)
\ee
where
$A_0$ is given in \rf{A01} and
\be{A2}
A_2=\frac13(\lb_3-e^{i\frac{\pi}{3}}\lb_2-
e^{-i\frac{\pi}{3}}\lb_1)
\ee
for the upper sign in \rf{Kqp}.
It follows from \rf{Psi2pq} that the vectors $\Psi$ have
the same permutation symmetry as in the previous case.
Hence, this is a set of symmetric states also.
The same conclusion takes place for
the lower sign in \rf{Kqp}.
Note that $A_2$ defined by \rf{A2} cannot be real.
Therefore the vectors \rf{Psie3} are not a particular case of
the set \rf{Psi2pq}.
Note also that the parameters $(\lb_1,\lb_2,\lb_3)$,
defining the vectors \rf{Psi2pq}, should be such that
$\lb_1^2+\lb_2^2+\lb_3^2=1$.
Otherwise the normalization condition of the vectors is
violated.

Thus we have proven that for $N=3$
any set of states satisfying Theorem \ref{th3} is a
set of symmetric states.
In Section \ref{N4} we shall show that for $N>3$ this theorem
describes a wider set of states than the set of symmetric states.

Before illustrating Theorem \ref{Th1}
 we will construct a general set of
 three vectors normalized to unity and
described by a Hermitian positive definite
$3\times 3$ matrix $\Psi$.
This is possible if instead of $U^\dg_3$ given in \rf{U3}
we will use $U_3$
so that now $\Psi=U_3E_0U_3^\dg$.
The normalization condition for this set
 is given  by the same Eq. \rf{NC} where
the matrix of the main determinant is replaced by its
transposed form, i.e. $u_{k,j}\to u_{j,k}$.
Since such a
replacement does not affect the main determinant, the
system remains to be overfull and one can express, for
instance, $\lb_1^2$ and $\lb_2^2$ in terms of $\lb_3^2$
\ba{lb123a}
\lb_1^2&=&\frac{1-3|u_{2,2}|^2+\lb_3^2(|u_{2,2}|^2-|u_{3,2}|^2)}%
{|u_{1,2}|^2-|u_{2,2}|^2}\\ \label{lb123b}
\lb_2^2&=&3-\lb_1^2-\lb_3^2\,.
\ea
Thus, we have obtained a four-parameter set of matrices
$\Psi$ where three parameters
come from $SU_2$ group
 and one parameter is $\lb_3>0$.

 In particular, the choice $K_{xy}=0$ and $K_{z\a}=\exp(i\pi/4)$
selects from this set a one parameter ($\lb_3$) subset
\be{1sub3}
\Psi=\frac16\left(
\begin{array}{lll}
6+\lb_+ &-i\sqrt{3}\lb_- & (1-i)\lb_+\\
i\sqrt{3}\lb_- &3(2+\lb_+)&(1+i)\sqrt{3}\lb_-\\
(1+i)\lb_+&(1-i)\sqrt{3}\lb_-&2(3+\lb_+)
\end{array}
\right)
\ee
where we have abbreviated
$\lb_+=\lb_2+\lb_3-2$, $\lb_-=\lb_2-\lb_3$.
The eigenvalues of $\Psi$ are
$\lb_1=1$, $\lb_2=\sqrt{2-\lb_3^2}$ and $\lb_3$.

As a numerical example we choose
 $\lb_3^2=3/2$.
 The optimal point on the sphere
 is found using Corollary \ref{Cr1},
  $(t_1,t_2)=(0.919,0.992)$,
 which corresponds to the weights
 $(x_1, x_2,x_3)=(0.67,0.43,0.99)$ and the optimal mean
 efficiency $P_M=0.535$.
The mean efficiency at the symmetric point
$t_0=(\arccos\frac{1}{\sqrt{3}},\pi/4)$,
 $P(t_0)=\lb_2^2=1/2<P_M$, is not optimal.

\subsection{$N=4$\label{N4}}

We start with an orthonormal set of $4$-vectors where one
of the vectors has equal coordinates
\be{U4}
U_4=\frac12\left(
\begin{array}{cccc}
0        & -1  &-\sqrt{2} &1\\
-\sqrt{2}& 1         & 0  & 1 \\
0        &  -1 & \sqrt{2} & 1\\
\sqrt{2} &   1     &    0 & 1
\end{array}
\right).
\ee
In general one can apply a $8$-parameter transformation
from the group $SU_3$ to the vectors $(u_1,u_2,u_3)$ to get
a general orthonormal set having a vector with equal
coordinates. For the sake of simplicity we will use $SU_2$
group once again and apply it to the vectors
$u_1$ and $u_2$ in a similar way as it was described
in the previous section.
In such a way one obtains the following elements of
matrix 4$\Psi$:
\begin{equation}\label{4p4}
\begin{array}{l}
\psi_{1,1}=2\lb_3+\lb_4+\lb_1|K_{xy}|^2+\lb_2|K_{z\a}|^2\,,
\\
\psi_{2,2}=\lb_4+\lb_2|i\sqrt{2}K_{xy}+K_{z\a}|^2
+\lb_1|iK_{xy}-\sqrt{2}K_{z\a}|^2\,,
\\
\psi_{3,3}=\psi_{1,1}\,,
\\
\psi_{4,4}=\psi_{2,2}+4\sqrt{2}(\lb_1-\lb_2)\im(K_{xy}^*K_{z\a})\,,
\\
\psi_{1,2}=-\psi_{1,1}+2(\lb_3+\lb_4)-
i\sqrt{2}(\lb_1-\lb_2)K_{xy}^*K_{z\a}\,,
\\
\psi_{1,3}=\psi_{1,1}-4\lb_3\,,
\\
\psi_{1,4}=4(\lb_3+\lb_4)-2\psi_{1,1}-\psi_{1,2} \,,
\\
\psi_{2,3}=\psi_{1,2}^*\,,
\\
\psi_{2,4}=\psi_{1,1}-\psi_{2,2}+\psi_{1,3}+\psi_{1,4}^*-\psi_{1,2}^*\,,
\\
\psi_{3,4}=\psi_{1,4}\,.
\end{array}
\end{equation}
One can easily check that after
this transformation system \rf{NC} at $N=4$ is overfull and
has a solution
$\lb_1^2=\lb_2^2=\lb_3^2=\frac13(4-\lb_4^2)$
useless for our purposes since it corresponds to symmetric
states.
To get another solution we will assume
$\lb_i^2\ne\lb_j^2$ for $i\ne j$ and solve the system
\rf{NC} with respect to $SU_2$ group parameters
thus obtaining
\ba{Kyz2}
|K_{xy}|^2=\frac{\lb_1^2-\lb_3^2}{\lb_1^2-\lb_2^2}\,,
\quad
|K_{z\a}|^2=\frac{\lb_3^2-\lb_2^2}{\lb_1^2-\lb_2^2}\,,
\\
K_{xy}=\pm\sqrt{\frac{\lb_3^2-\lb_1^2}{\lb_2^2-\lb_3^2}}K_{z\a}\,,
\label{Kyz22}
\ea
With this set of parameters
the vectors $\Psi=\Psi^\dg$ have unit norm
for arbitrary $\lb_j$
provided
$\sum_{j=1}^4\lb_j^2=4$.
For the upper sign in \rf{Kyz22}
the basic elements of the matrix $4\Psi$ given in \rf{4p4}
read
\be{4p4e}
\begin{array}{l}
\psi_{1,1}=\lb_1+\lb_4+2\lb_3+\frac{\lb_2^2-\lb_3^2}{\lb_1+\lb_2}\,,
\\
\psi_{2,2}=\psi_{1,1}+2\frac{(\lb_1-\lb_3)(\lb_2-\lb_3)}{\lb_1+\lb_2}\,,
\\
\psi_{4,4}=\psi_{2,2}\,,
\\
\psi_{1,2}=2(\lb_3+\lb_4)-\psi_{1,1}+
\frac{2iB_1}{\lb_1+\lb_2}\,,
\end{array}
\ee
where
\be{B1}
B_1=\frac{\sqrt{(\lb_3^2-\lb_2^2)(\lb_1^2-\lb_3^2)}}{\sqrt{2}}\,.
\ee
Thus we have obtained a 3-parameter
($\lb_2,\lb_3,\lb_4$) set of the vectors $\Psi$.

Note that Eq. \rf{Kyz2}
imposes restrictions on possible values of $\lb_j^2$.
In particular, the following inequalities:
\be{inequ}
\lb_4^2<\lb_2^2<\lb_3^2<\lb_1^2
\ee
 should hold.
Here we imposed the condition that $\lb_4^2$ is the lowest
eigenvalue of $\Psi^2$.
Since under the last restriction the eigenvector of $\Psi^2$
corresponding to the lowest eigenvalue has equal
coordinates, we can apply
Theorem \ref{th3}.
According to this theorem the optimal mean efficiency
coincides with $\lb_4^2$,
$P_M=\lb_4^2$,
provided the parameters are ordered according to \rf{inequ}
and $\lb_2^2+\lb_3^2+\lb_4^2<4$.

Finally, to show that the vectors from this set are neither
symmetric states nor their unitary rotated form we give
matrix $\Psi^2$
\be{psi24}
\Psi^2=\left(
\begin{array}{llll}
1&B_2^*&B_3&B_2\\
B_2&1&B_2&B_3+iB_1\\
B_3&B_2^*&1&B_2\\
B_2^*&B_3-iB_1&B_2^*&1
\end{array}
\right)
\ee
where $B_1$ in given in \rf{B1},
$B_2=\frac12(2-\lb_1^2-\lb_2^2-iB_1)$
and
$B_3=1-\lb_3^2$.
In the next section we will show that
a matrix $\Psi^2$ of the form \rf{psi24}
cannot correspond
to symmetric states.

\subsection{Symmetric states\label{SymS}}

In this section we prove that
the mean efficiency for the equiprobable
so called symmetric
states (see e.g. \cite{8} for some discussion and literature
overview)
is maximal at the symmetric point on the unit sphere.

As shown in \cite{8} symmetric states $\psi_k$ may be defined in
terms of any orthonormal basis $\{e_k\}$, $k=0,\ldots,N-1$
$\la e_k|e_j\ra=\d_{kj}$ in $\cH$ as follows
\be{SyS}
\psi_k=\sum_{j=0}^{N-1}c_je^{2\pi ijk/N}e_j\,,\quad
k=0,\ldots,N-1,
\ee
where coefficients $c_k$ may be arbitrary complex numbers
provided
\be{norm}
\sum_{k=0}^{N-1}|c_k|^2=1\,.
\ee

In what follows we choose a representation where
$(e_0,\ldots,e_{N-1})=I_N$.
Then
using the fact that the additional phase factors in \rf{SyS}
are either $N$th roots of unity or their integer powers and
therefore they satisfy the following identity
\be{}
\sum_{k=0}^{N-1}e^{2 \pi i (j-j')k/N}=N\d_{jj'}\,,
\quad j,j'=0,1,\ldots
\ee
one easily sees that
\be{PP}
\Psi\Psi^\dg=N\mbox{diag}(|c_0|^2,\ldots,|c_{N-1}|^2)
\ee
where as above $\Psi=(\psi_0,\ldots,\psi_{N-1})$.
By the same reason and in view of condition
\rf{norm}
matrix $\Psi^\dg\Psi$ has the form
\be{circ}
\Psi^\dg\Psi\equiv A=\left(
\begin{array}{lllll}
A_0 & A_{N-1} & A_{N-2} & \cdots &A_{1}\\[1em]
 A_{1} & A_0 & A_{N-1} & \cdots & A_{2}\\[1em]
 A_2 & A_{1} & A_0&  \cdots & A_{3} \\[1em]
 A_3 & A_{2} & A_{1} & \cdots &A_{4} \\
\vdots &\vdots &\vdots & \ddots & \vdots \\
A_{N-1} & A_{N-2} & A_{N-3} & \cdots &A_0
\end{array}
\right)
\ee
where $A_0=1$ and
\be{}
A_k=\sum_{j=0}^{N-1}|c_j|^2e^{2 \pi ijk/N}\,, \quad
k=1,\ldots,N-1\,.
\ee
A
matrix of the form \rf{circ} is called {\it circulant}
\cite{MM}.
For any complex $A_k$ it has eigenvalues \cite{MM}
$F(\e^{k+1})$, $k=0,\ldots,N-1$ where
\be{Fx}
F(x)=\sum_{k=0}^{N-1}A_kx^k\,,\quad
\e=e^{2 \pi i/N}
\ee
and eigenvectors
\be{vk}
v_{k-1}=\frac{1}{\sqrt{N}}\sum_{j=1}^N\e^{(N-k)j}e_{j-1}\,,\quad
k=1,\ldots,N\,,
\ee
\be{}
Av_k=F(\e^{k+1})v_k\,,\quad k=0,\ldots,N-1\,.
\ee
From \rf{vk} it follows that if $v_{k,j}$ are the
coordinates of the
vectors $v_k$ then $|v_{k,j}|=N^{-1/2}$ for $k,j=0,\ldots,N-1$.
Using Theorem \ref{th3}, the fact that matrices $\Psi\Psi^\dg$
and $\Psi^\dg\Psi$ have the same set of eigenvalues
and formula \rf{PP}
we obtain for $P_M$
 just the value reported by Chefles and
Barnett \cite{8}
\be{CBSS}
P_M=N\times\min(|c_0|^2,\ldots,|c_{N-1}|^2)\,.
\ee

Note that the entries of matrix \rf{circ} are inner
products of the vectors $\psi_j$, i.e.
 $A=(\la\psi_i|\psi_j\ra)$.
 Therefore they are independent on a unitary
 rotation of the set $\Psi$.
 By this reason, for any set of
symmetric states $\Psi$, matrix of their inner products has
 always form \rf{circ}.
 The opposite statement, evidently, is also true.
 If for a
 set of states $\Psi$ the matrix $\Psi^\dg\Psi$ has the
 form \rf{circ}, then the states are either symmetric states or their
 unitary rotated form.
 This follows from the fact that one
 of the square roots of matrix \rf{circ} is a
 Hermitian matrix of symmetric states.
If for a Hermitian matrix $B^2=(B^2)^\dg$ a Hermitian
matrix $B=B^\dg$ is given, then
 any other Hermitian square root of $B^2$,
$\tilde B=\tilde B^\dg$, $\tilde B^2=B^2$,
  is
 defined up to a unitary matrix  $V=(V^\dg)^{-1}$,
$\tilde B=VB$ such that $VB=BV^\dg$.

The entries of a matrix given in \rf{NN1} and
\rf{NN2} correspond to a set of symmetric states.
This means
that any set of states satisfying Theorem \ref{th4}, i.e. a
set, for which one of the eigenvalues of the matrix $\Psi^\dg\Psi$
is
$N-1$ fold degenerate,
is either a set of symmetric states or its
unitary rotated form.
From this viewpoint Theorem \ref{th4}
gives another recipe how one can identify such a set
when it is presented by a non-Hermitian matrix.
From the other hand matrix \rf{psi24} does not have
this form meaning that the states $\Psi$ obtained in
Section \ref{N4} are not symmetric.

Another remarkable property of matrix \rf{circ} we
would like to emphasize is the following.
Assume we are given vectors collected as the columns to matrix
\rf{circ}, i.e. we put
$\Psi=(\psi_1,\ldots,\psi_N)=A$.
In this case, evidently,
 $\psi_{j+1}=P\psi_j$,
$j=1,\ldots,N-1$, $\psi_1=P\psi_N$ where $P$ displaces
every coordinate of a vector to the next position and
places the last coordinate in place of the first one.
Such
an operator is unitary and, hence, according to  a
definition of symmetric states
(see e.g. \cite{8})
 the states \rf{circ} are symmetric.
 Using Theorem \ref{th3}
 we can obtain another
(but equivalent to \rf{CBSS})
  expression for the optimal mean
 efficiency for symmetric states.
 It may be useful when
 symmetric states are given in a form different from that
 used by Chefles and Barnett \cite{8}.

This possibility is based on the property of matrix
\rf{circ}
to be normal, i.e. $AA^\dg=A^\dg A$ which can be checked by
a direct calculation.
Therefore matrices $A$, $A^\dg$ and $A^\dg A$ may have the same
set of eigenvectors given in \rf{vk}.
Absolute values of coordinates of any eigenvector coincide
and we can apply Theorem \ref{th3} to find the optimal
mean efficiency.
According to Corollary
\ref{prop} for the optimization procedure we can replace
$\Psi$ by positive definite Hermitian matrix
$\Psi=U^\dg E_0U$ where diagonal matrix $E_0$ contains
absolute values of matrix $A$ eigenvalues as non-zero entries,
$E_{0,k}=|F(\e^{k+1})|$, $k=0,1,\ldots,N-1$
where $F(x)$ and $\e$ are given in \rf{Fx}.
To apply Theorem \ref{th3}
the vectors $\Psi=A$ should have unit norm.
This condition
is satisfied if $\sum_{j=0}^{N-1}|A_j|^2=1$.
Now using Corollary \ref{Cr1} we
conclude that
$P_M=\mbox{min}(|E_{0,0}|^2,\ldots,|E_{0,N-1}|^2)$.
Parameters $A_k$ should be chosen such that $E_{0,k}\ne0$,
$k=0,\ldots,N-1$. Otherwise the set $\Psi$ becomes linearly
dependent.

\section{POVM and Neumark's extension}

In the previous section we formulated conditions to be
imposed on the set of the vectors $\Psi$ when, from the one
hand, the mean efficiency takes its maximal value
at the symmetric point
and from
the other hand there exists a Hermitian positive
semidefinite
 matrix $\wt\Pi_x$ satisfying \rf{POVMx}.
 Another important property we established is that
 although $\wt\Pi_x$ is an $N\times N$ matrix its
  rank may vary from $1$ till $N-1$ so that if we
  associate a set of the vectors with the columns of this
  matrix, this set is necessarily linearly dependent with a
  linearly independent subset spanning the space of
  dimension $N_a$ varying from $1$ till $N-1$.
 Below we will discuss the so called Neumark's extension which,
 in particular,
 consists in considering an additional space called
 {\em ancilla space} related just with the set of the vectors collected
 to $\wt\Pi_x$.
 In particular, the dimension of the ancilla space is
 $N_a=\mbox{\rm rank}\,\wt\Pi_x$.

 The notion of an ancilla first introduced in quantum
information theory \cite{Hel}
permits one to reduce generalized measurements to conventional
von Neumann measurements in a higher dimensional space
 \cite{GM}.
Usually this extended space is
considered as
the direct product
of spaces of the system to be
measured with another
(auxiliary) known system called the ancilla
(see e.g. \cite{Peres-FF}).
But as is stressed in \cite{Chen}
it may have a wider sense
as any extra degrees of freedom related with the Neumark's
extension theorem.
Since there are two ways to
extend the initial Hilbert space, there are two different
realizations of the extended Hilbert space:
 (i) the
extended space is the tensor product of the initial
(i.e. system) space and
the ancilla space (TPE) and (ii) the extended space is the
direct sum of the initial space and the ancilla space (DSE)
\cite{Chen}.
As is stressed in \cite{Chen} DSE is much more
economical from the experimental viewpoint since in this case the
dimension of the extended space is much less than
the corresponding dimension for TPE.
In particular, the authors \cite{Chen} claim that to
unambiguously discriminate among $N$ given linearly
independent non-orthogonal states
by DSE
 an $(N-1)$-dimensional ancilla space is sufficient
 whereas for TPE $N^2$ additional dimensions are necessary,
 where $N$ is just the space dimension of the main system, i.e.
 the number of the given linearly independent vectors
 collected to $\Psi$.
From the other hand, as it is pointed out by Preskill
\cite{GM}, the tensor product extension is much more
transparent from physical viewpoint since additional
dimensions can always be interpreted as extra degrees
of freedom of the extended system
(e.g. internal degrees of freedom like spin or external degrees
of freedom corresponding to an interaction of the system
 with a reservoir).

 As is stressed in Section \ref{OP}, $(N-1)$-dimensional
 ancilla space is the maximal space appearing
 as result of the optimization procedure and the actual ancilla
 space
 dimension varies from $1$ till $N-1$.
Thus for $N>2$ we give a stronger limit to the minimal ancilla
 space dimension for the DSE-case than the one found in \cite{Chen}.
 To the best of the
 author's knowledge in the current literature there are no
 explicit constructions of POVMs accepting
 ancilla spaces of such low dimension as $1$ for
 $N>2$
 \footnote{For N=2 there is only one possibility,
$N-1=1$.
Explicit examples of the one-dimensional ancilla space
for $N=2$ are given in \cite{Chen} and earlier in
S. Franke-Arnold, E. Andersson, S. M. Barnett and S. Stenholm,
Phys. Rev. A {\bf 63}, 052301 (2001).}.
 The next section is
 just devoted to fill in this gap. We will present a
 general construction of a simple form for a POVM with the
 ancilla space dimension in the range between $1$ and $N$.

 Another point we would like to emphasize is the following.
 According to the necessary condition \cite{8} for the optimization of
 the mean efficiency the highest singular number of matrix
 $\Xi_x$ should be equal to $1$.
 Just this property leads to
 a linear dependence between the vectors $\wt\Pi_x$. If
 after the optimization one can sacrifice the very optimal
 POVM in favor of a bit less optimal POVM where all the
 vectors collected to matrix $\Xi_x$ are scaled by a factor a
 bit less than $1$ (real scaling factor should be chosen
 in agreement with how much from the optimal efficiency one
 is able to sacrifice), then the rank of $\wt\Pi_x$ becomes equal
 to $N$ and the ancilla space acquires the same dimension as
 the system space ($N_a=N$).
 In such a case, as we show below by an
 explicit construction of the corresponding Neumark's
 extension, one may enjoy an advantage of the direct
sum extension
of two copies of the same $N$-dimensional space
being presented
in the form of
  the tensor product of the
two-dimensional ancilla space and
 $N$-dimensional system space
 with a clear physical meaning
of the ancilla space
 as an additional (e.g. internal) degree
 of freedom of an extended system.
 Thus, similar to DSE-case, we indicate on a
stronger limit of the ancilla space dimension for
 the TPE-case which is $2$ instead of $N+1$ found in
 \cite{Chen}. The dimensionality of the enlarged
 (system+ancilla) space in this case is $2N$
 versus $N(N+1)$ indicated in \cite{Chen}.

Below we will consider only the vectors $\Xi_x$ and
 to simplify notations will omit subscript $x$.

\subsection{POVM}

Let the vectors $\xi_i\in\cH$, $i=1,\ldots,N$, collected to
the matrix $\Xi$, be given.
Denote as $\tXi$ matrix with
unknown linearly independent vectors
$\txi_j\in\cH$, $j=1,\ldots,N_a\le N$ such that
 the identity decomposition
\be{POVM}
\Xi\Xi^\dg+\tXi\tXi^\dg=I_N
\ee
holds.
Since by assumption the vectors $\Xi$ are linearly
independent $\mbox{rank}\Xi=N$ but
$\mbox{rank}\tXi=N_a\le N$ and
 $\Xi\Xi^\dg$ is positive definite while
 $\tXi\tXi^\dg$ may be positive both definite and
 semidefinite.
In both cases matrix $(I_N-\Xi\Xi^\dg)^{1/2}$ is
well defined and Hermitian.
Therefore from \rf{POVM} it follows
\be{ttpsi}
\tXi=(I_N-\Xi\Xi^\dg)^{1/2}\wt V\,,\quad
\wt V^{-1}=\wt V^{\dg}
\ee
where $\wt V$ is an $N\times N$ arbitrary unitary matrix.
In particular,
 for $N_a=N$ one may choose
$\wt V=(\Xi\Xi^\dg)^{-1/2}\Xi$ leading to
\be{tXi}
\tXi=(I_N-\Xi\Xi^\dg)^{1/2}(\Xi\Xi^\dg)^{-1/2}\Xi
=[(\Xi\Xi^\dg)^{-1}-I_N]^{1/2}\Xi\,.
\ee

Note that the $N\times N$ matrix $\tXi$ \rf{tXi}
contains $N$ linearly independent
 vectors as columns only if $N_a=N$.
If $N_a<N$
among $N$ columns of this matrix only $N_a$ columns
are linearly independent.
Since we want this matrix to contain only linearly independent
columns it should be rectangular of dimension
$N\times N_a$ in this case.
Moreover, there is no need to rotate all $N$ vectors by
transformation $\wt V$. It is sufficient to use
an arbitrary
$N_a\times N_a$ unitary rotation instead of $\wt V$.
Below we explicitly construct a matrix
$\tXi$ having this property.

For $N_a<N$
denote as $U$ unitary matrix bringing $\Xi\Xi^\dg$ to a diagonal
form
\be{}
\Xi\Xi^\dg U=U S_d
\ee
where $S_d$ is a diagonal matrix,
in which the highest $(N-N_a)$ fold degenerate eigenvalue,
which is equal to $1$, occupies the last
$(N-N_a)$ positions.
Then from \rf{POVM} one finds
\be{}
\tXi\tXi^\dg=U(I_N-S_d)U^\dg\,.
\ee
It is clear that by construction
 only the first $N_a$
columns of matrix $U(I_N-S_d)^{1/2}$ are nonzero.
Therefore, the solution we need is
obtained if $\tXi$ is composed of all non-zero columns of
 $U(I_N-S_d)^{1/2}$,
which we denote $\tXi_1$,
  times an arbitrary unitary $N_a\times N_a$ matrix
 $V$, i.e.
 \be{tpsit}
\tXi=\tXi_1V\,,\quad
 V^{-1}=V^{\dg}\,.
 \ee
Note that all $N_a$ columns of
$N\times N_a$ matrix $\wt\Xi$ \rf{tpsit} are linearly
independent by construction.

\subsection{Neumark's extension}

Neumark \cite{Neum} proved a very general statement
concerning a representation of an additive operator-valued
function in a unitary space in terms of orthogonal
spectral functions in higher spaces. In the context of our
problem it, in particular, means that any non-orthogonal
identity decomposition
in the space $\cH_N$
 may be presented as a projection from a space of a higher
 dimension of an orthogonal identity decomposition
 (see e.g. \cite{GM}).
  Although there exists a number of
 constructions extending a given non-orthogonal identity
 decomposition to an orthogonal one in a
 higher dimensional space
 (see e.g. \cite{GM,Chen,He}) they are based on either the
 tensor product extension \cite{GM} or are not suitable for
 establishing an isomorphism between Hilbert spaces
$\cH_N\oplus\cH_N$ and $\cH_2\otimes\cH_N$.
 Therefore in this section we
 give an explicit realization of the Neumark's theorem
  suitable for this purpose.

Let us consider $(N+N_a)\times(N+N_a)$ matrix
($N_a\le N$)
\be{U}
\cU=\left(
\begin{array}{cc}
\Xi &\tXi\\ Z &Y
\end{array}
\right).
\ee
As it is noted by Preskill \cite{GM} the rows
$(z_{k,1},\ldots,z_{k,N},y_{k,1},\ldots,y_{k,N_a})$,
$k=N+1,\ldots,N+N_a$ of matrix \rf{U} should be orthogonal to
all its rows
$(\xi_{j,1},\ldots,\xi_{j,N},\wt\xi_{j,1},\ldots,\wt\xi_{j,N_a})$,
$j=1,\ldots,N$
where $z_{k,j}$, $y_{k,j}$, $\xi_{k,j}$ and $\wt\xi_{k,j}$ are
entries of blocks $Z$, $Y$, $\Xi$ and $\tXi$ of $\cU$
respectively.
Below we find
explicit expressions for $Z$ and $Y$ in terms of the given
$\Xi$ and $\tXi$
such that $\cU$ is
unitary
\be{cUI}
\cU^\dg\cU=I_{N+N_a}
\ee
(or equivalently $\cU\cU^\dg=I_{N+N_a}$).

Condition \rf{cUI} is equivalent to
the following set of equations for $Z$ and $Y$
\ba{XYeq}
\Xi^\dg\Xi+Z^\dg Z=I_N\,, \label{XY1}\\
\tXi^\dg\tXi+Y^\dg Y=I_{N_a}\,,\label{XY2}\\
\Xi^\dg\tXi+Z^\dg Y=0\label{XY5}\,.
\ea
Now we note that the condition $\cU\cU^\dg=I_{N+N_a}$  is
an implication of \rf{cUI}.
This, in particular, means that the POVM condition \rf{POVM}
follows from  \rf{XY1}-\rf{XY5}.
Therefore we can replace Eq. \rf{XY2} by Eq. \rf{POVM}.
From \rf{XY1} we find matrix $Z$.
For that we multiply \rf{POVM} from the left by
unitary matrix $\Phi^\dg$ and from the right by $\Phi$
where $\Phi=(\Xi\Xi^\dg)^{-1/2}\Xi$
which gives
$\Xi^\dg\Xi+\Phi^\dg\tXi\tXi^\dg\Phi=I_N$.
Comparing this equation with \rf{XY1} we conclude that
$Z^\dg Z=\Phi^\dg\tXi\tXi^\dg\Phi$ and, hence,
\be{X}
 Z=V\tXi^\dg\Phi=V\tXi^\dg(\Xi\Xi^\dg)^{-1/2}\Xi\,,
 \quad V^{-1}=V^{\dg}
 \ee
 where $V$ is $N_a\times N_a$ arbitrary unitary matrix.
Note that $ZZ^\dg=V\tXi^\dg\tXi V^{\dg}$
is $N_a\times N_a$
overlap matrix of the unitary rotated linearly independent
by construction
vectors $\txi_i$,
$i=1,\ldots,N_a$ collected as  columns to $\tXi$ and,
therefore,
 $ZZ^\dg$ is non-singular.
 Using this fact we find from \rf{XY5}
\be{Y}
Y=-V(\tXi^\dg\tXi)^{-1}\tXi^\dg(\Xi\Xi^\dg)^{1/2}\tXi\,.
\ee
Note that for $Y$ given in \rf{Y} condition \rf{XY2} is
automatically satisfied which may be checked by a direct
calculation.

It is important to note that if a non-singular $\tXi$ given
in \rf{tXi} is used in \rf{U}, in which case $N_a=N$,
Eqs. \rf{X} and \rf{Y} assume simpler form
\ba{XYZ}
Z&=&V\Phi^\dg\tXi \label{Xz}\,,\\
Y&=&-V\Phi^\dg\Xi \label{Yz}\,.
\ea
These equations suggest the choice $V=\Phi$ leading to the
simplest expressions
 for $Z$ and $Y$
\be{XN}
Z=\tXi\,, \quad Y=-\Xi\,,
\ee
and the following expression for $\cU$
\footnote{For a particular case of $2\times2$ matrices
see also
  U. G\"{u}nther and B. F. Samsonov, Phys. Rev. Lett. {\bf 101}, 230404 (2008).}
\be{cU}
\cU=\sigma_z\otimes\Xi+\sigma_x\otimes\tXi\,,
\ee
where $\sigma_{x,z}$ are the usual Pauli matrices.
Since all Hilbert spaces of the same dimension are
isomorphic to each other, $2N$-dimensional Hilbert space,
which is the direct sum of two $N$-dimensional spaces,
$\cH_N\oplus\cH_N$, is isomorphic to
$\cH_2\otimes\cH_N$.
Relation \rf{cU} may be used to map the operator $\cU$
presented by matrix \rf{U} in the space $\cH_N\oplus\cH_N$
to the space $\cH_2\otimes\cH_N$.

\section{Conclusion}

The main feature of our approach to optimization of the
mean efficiency for discriminating among $N$ given
linearly independent non-orthogonal states from methods
usually used by previous authors consists in choosing a
different form for the constraint to be imposed on
optimization parameters to assure the existence of a POVM and,
hence, the necessary probabilistic interpretation.
The form of the constraint, we have used,  permitted us
to reduce the conditioned optimization problem for
the mean efficiency to a problem of
finding an unconditioned maximum of a
function defined on an $N$-ellipsoid (Theorem \ref{Th1})
for the states given with different probabilities and on a
unit $N$-sphere for equiprobable states (Corollary \ref{Cr1}).
Using Theorem \ref{Th1} we established the invariance of
both the weight matrix and the optimal mean efficiency with
respect to unitary rotations of the vector set.
Therefore
for any vector set $\Psi$ the optimization procedure may be
realized for its equivalent Hermitian form,
$\Psi=\Psi^\dg$ with $\Psi$ being a positive definite
matrix.
Using this fact for equiprobable states
 we succeeded to formulate a criterion when
the optimal point is the symmetric point on the $N$-sphere.
This selects a set of matrices for which a numerical
optimization becomes unnecessary.
By the symmetric point on the sphere we mean a point with
equal values of Cartesian coordinates when the sphere is
centered at the origin.
We have shown that for $N=2,3$ only symmetric states
satisfy this criterion but starting from $N=4$ the indicated set
becomes wider than the
set of symmetric states.
The whole family of states satisfying
this criterion is still unspecified.

We have also found
a subset of $N$ symmetric states which may be
considered as the simplest generalization of two states
since the expression for the optimal mean efficiency
for two states is a particular
case of a more general formula valid for the subset.
The vectors from this subset are characterized by the following property.
If $\Psi$ is a matrix where the state vectors
are collected as columns then $\Psi^\dg\Psi$
has only
two distinct eigenvalues one of which is $N-1$ fold
degenerate.

Our approach is illustrated by
 examples with $N=2,3,4$ nonorthogonal states.

Finally we presented our constructions of POVM and
Neumark's extension.
We indicated on an explicit procedure how
to construct a POVM for the case when the dimension of the
ancilla space varies from $1$ till $N$.
 As to Neumark's
extension we presented a formula
for the unitary matrix realizing the corresponding
orthogonal identity decomposition
 suitable for its use
both in the Hilbert space
$\cH_N\oplus\cH_N$ and $\cH_2\otimes\cH_N$.

\section*{Acknowledgments}
The author would like to thank the referee for valuable
comments.
 He is also grateful to U. G\"unther for useful discussions.
This work is partially supported by the grants
RFBR-09-02-00009a
and SS-871.2008.2.

\end{document}